\documentclass{optica-article}
\journal{oe}
\articletype{Research Article}
\usepackage{upgreek}
\usepackage{svg}
\usepackage{braket}

\begin{document}

\title{Ultra-high Q alumina optical microresonators in the UV and blue bands}

\author{Chengxing He\authormark{1,3}, Yubo Wang\authormark{1}, Carlo Waldfried\authormark{2}, Guangcanlan Yang\authormark{1},  Jun-Fei Zheng\authormark{2}, Shu Hu\authormark{3} and Hong X. Tang\authormark{1,*}}

\address{

\authormark{1}{Department of Electrical Engineering, Yale University, New Haven, CT 06520, USA} \\
\authormark{2}{Entegris Inc., Billerica, MA 01821, USA} \\
\authormark{3}{Energy Sciences Institute, Yale University, West Haven, CT 06516, USA}
}

\email{\authormark{*}hong.tang@yale.edu} 



\begin{abstract}
UV and visible photonics enable applications ranging from spectroscopic sensing to communication and quantum information processing. Photonics structures in these wavelength regimes, however, tend to experience higher loss than their IR counterpart. Particularly in the near-UV band, on-chip optical microresonators have not yet achieved a quality factor beyond 1 million. % Alumina, known for its low optical absorption and large bandgap, is an attractive photonic material for short wavelength operations. 
Here, we report ultra-low-loss photonic waveguides and resonators patterned from alumina thin films prepared by a highly scalable atomic layer deposition process. %With high confinement ring resonators %whose footprint measures smaller than 1mm $\times$ 1mm, 
We demonstrate ultra high Q factor of 1.5$\,\times\,$10$^6$ at 390\,nm, %(optical loss of $\sim$\,0.84\,dB/cm)
a record value at UV bands, and 1.9$\,\times\,$10$^6$ at 488.5\,nm. %(optical loss of $\sim$0.51\,dB/cm). Compared to visible photonics that employs Si$_3$N$_4$ as the waveguide core material, the alumina photonics platform shows clear advantages in the UV-blue bands by offering lower optical loss and a smaller footprint. The thin film manufacturing process is foundry compatible, thus establishing reliable UV and blue band photonics based on alumina thin films.
\end{abstract}

%%%%%%%%%%%%%%%%%%%%%%%%%%  body  %%%%%%%%%%%%%%%%%%%%%%%%%%

\section{Introduction}

UV and visible band integrated photonics has witnessed rapid progress in recent years. Applications such as atomic clocks\cite{AMOARC}, biochemical sensing\cite{biosensing,biosensing2}, visible light communications\cite{LiFi}, quantum sensing \cite{atomsensing,Rydsensing}, quantum information processing based on trapped ions\cite{mehta2020integrated,niffenegger2020integrated} and atoms\cite{6MAMO,AMO780}, all call for UV and blue band photonic integrated circuits with high scalability and low loss. Yet, low-loss photonics at short wavelengths remains difficult to achieve as material absorption dramatically increases when photon energy approaches the bandgap of a material, and Rayleigh scattering scales as $\lambda^{-4}$. One approach to reduce the loss at short wavelengths is to use very thin silicon nitride (Si$_3$N$_4$) waveguides cladded with low loss silica\cite{5M,6MAMO} so that the propagation mode is weakly confined. In this way, absorption inherent to the Si$_3$N$_4$ waveguide core can be diluted, while scattering loss induced by waveguide sidewall roughness is also reduced due to the reduced sidewall height. However, so far, devices employing Si$_3$N$_4$ as the waveguide core still show strong absorption in the UV and blue bands.\cite{hugerange,SiN410}   %However, these thin waveguide designs are prohibited from transmitting TM modes, and ultimately, the scattering loss induced by the overlap between the propagation mode and the top and bottom surface of the waveguide, as well as the absorption loss induced by Si$_3$N$_4$ core cannot be mitigated. 

Further reduction of waveguide loss at short wavelengths requires a new waveguide core material that demonstrates a large bandgap while still maintaining a higher refractive index than the low-loss cladding material, usually silica, to provide good confinement. One such candidate is AlN, which has a large bandgap of 6.2\,eV. In our previous work, we employed single-crystalline AlN as waveguide core\cite{AlN} %AlN has a large bandgap of 6.2 eV %and offers intrinsic $\chi^{(2)}$ and $\chi^{(3)}$ susceptibilities for nonlinear optics applications
and demonstrated a quality factor of 210\,k at 390\,nm, which was a significant advance for devices operating in the near-UV bands, but still below the state-of-the-art achieved by IR and near-visible optical resonators. %, especially considering their small footprints of just 30$\mu$m radius. 

An alternative to AlN for short-wavelength passive photonic integrated platforms is amorphous alumina. Recent progresses in deposition methods have greatly improved the quality of amorphous alumina films, which %see fewer defects than previous endeavors and 
now demonstrate a band gap that is comparable to bulk sapphire (7.0 - 8.3\,eV of ALD alumina \cite{alumina1,alumina2} vs. 8.8\,eV\cite{sapphire}). Many reports on amorphous alumina films deposited via either reactive sputtering or atomic layer deposition (ALD) also confirmed very low loss at short wavelengths (< 0.3\,dB/cm at 405 \,nm)\cite{aluwgd,aluwgd2,activealufilm}, and their compatibility with photonic platforms, either standalone \cite{activealu} or combined with other materials\cite{SiNAlu}. Recently, near-UV extended cavity diode lasers (ECDLs) were demonstrated by interfacing low-loss alumina waveguides with InGaN semiconductor amplifiers\cite{twente}. Thanks to the amorphous microstructure of these alumina films, the associated deposition process also has no requirements for the lattice structure of the substrate on which it is grown, thus relaxing requirements for substrate material. Furthermore, both ALD and reactive sputtering process are CMOS compatible, paving the way for CMOS integration with amorphous alumina-based photonics. 

In this letter, we leverage an industrial ALD process to grow alumina as the waveguide core. This highly scalable process is capable of providing uniform growth coverage to substrates over 20" in diameter and can coat hundreds of 4" wafers in a single batch. Because the absorption of UV and blue light is low in alumina, the propagation mode can be fully supported in the waveguide core, %thus reducing the footprint of resonators while 
minimizing the scattering loss at top and bottom surfaces of the alumina film. With new waveguide core material and corresponding design principles, our resonators demonstrated ultra high Q of 1.5$\,\times\,$10$^6$ at 390\,nm and 1.9$\,\times\,$10$^6$ at 488.5\,nm. Those are the highest quality factors reported at corresponding wavelengths for resonators featuring high confinement design, including previously demonstrated alumina resonator \cite{aluwgd,twente}. 

\section{Design and Simulation}
%Fig 1. shows the simulated modal profile of our resonators for different waveguide cross sections. Given the large bandgap of amorphous alumina, the absorption loss at wavelengths greater than 390\,nm should be dominated by scattering loss at rough surfaces of the alumina waveguide, particularly rough sidewalls formed during the etching process. 
We employ a shallow etch geometry to minimize sidewall scattering loss. The resonators are air-cladded to promote refractive index contrast with the alumina core, reducing the etch depth needed for confinement.% formed during the etch process%  where the sidewall of the ridge waveguide is just high enough to provide confinement for the propagation mode. 
The etch depth is optimized by simulating the radiation loss of waveguides subject to different etch depths in Lumerical. For ring resonators with 400\,$\muup$m radius, it was found that radiation loss can be suppressed to less than 0.06\,dB/cm at 488.5\,nm and less than 0.001\,dB/cm at 390\,nm when the etch depth is greater than 80\,nm out of 400\,nm thick alumina film. During fabrication, the etch depth is targeted at 100\,nm, deep enough so that the resonators are not radiation loss limited. %To further minimize scattering loss, 
The width of the waveguide is set to 4.5\,$\muup$m so that the outer sidewall provides most of the confinement for the propagation mode, and the overlapping between propagation mode, particularly TE00 mode, and the inner sidewall is minimal, further reducing scattering loss. It should be noted that this wide waveguide allows for propagation of multiple propagation modes, however, as Fig. 2 shows, by optimizing bus-to-resonator coupling, the coupling to the higher order modes can be greatly suppressed.

% \begin{figure}[h!]
% \centering\includegraphics[width=\textwidth]{modes.jpg}
% \label{modes}
% \caption{Simulated TE00 propagation modes at 390\,nm (left) and 488.5\,nm (right) in a 400\,$\muup$m radius ring resonator with 400\,nm height and 4.5\,$\muup$m width, at varying etch depths of (a) 50\,nm (b) 100\,nm and (c) 200\,nm. The mode confinement becomes increasingly better as etch depth increases, at the price of more overlapping between propagation modes and sidewall, which results in scattering loss.\question{This figure is highly repetitive and does not offer much content. You may want to condense these into Figure 3. Add panel (c)(d) there to include these simulations.} }
% \end{figure}
We utilized an under-coupled point-coupling design to reduce coupling-induced loss and probe the intrinsic alumina resonator loss. The coupling geometry between straight bus waveguide and ring resonator is optimized by varying the bus waveguide width and the gap between the bus waveguide and the ring resonator. We use a combination of simulations with FIMMWAVE software and experimental data to optimize the parameters. The optimal coupling condition for 390\,nm light is 0.65\,$\muup$m wide bus waveguide and 1\,$\muup$m gap between bus waveguide and ring resonator. For 488.5\,nm light, the optimum coupling condition is 0.75\,$\muup$m wide bus waveguide and 1.1\,$\muup$m gap. We also fabricated and measured microrings with smaller gaps, which provide stronger coupling.

 \begin{figure}[h!]
 \centering\includegraphics[width=\textwidth]{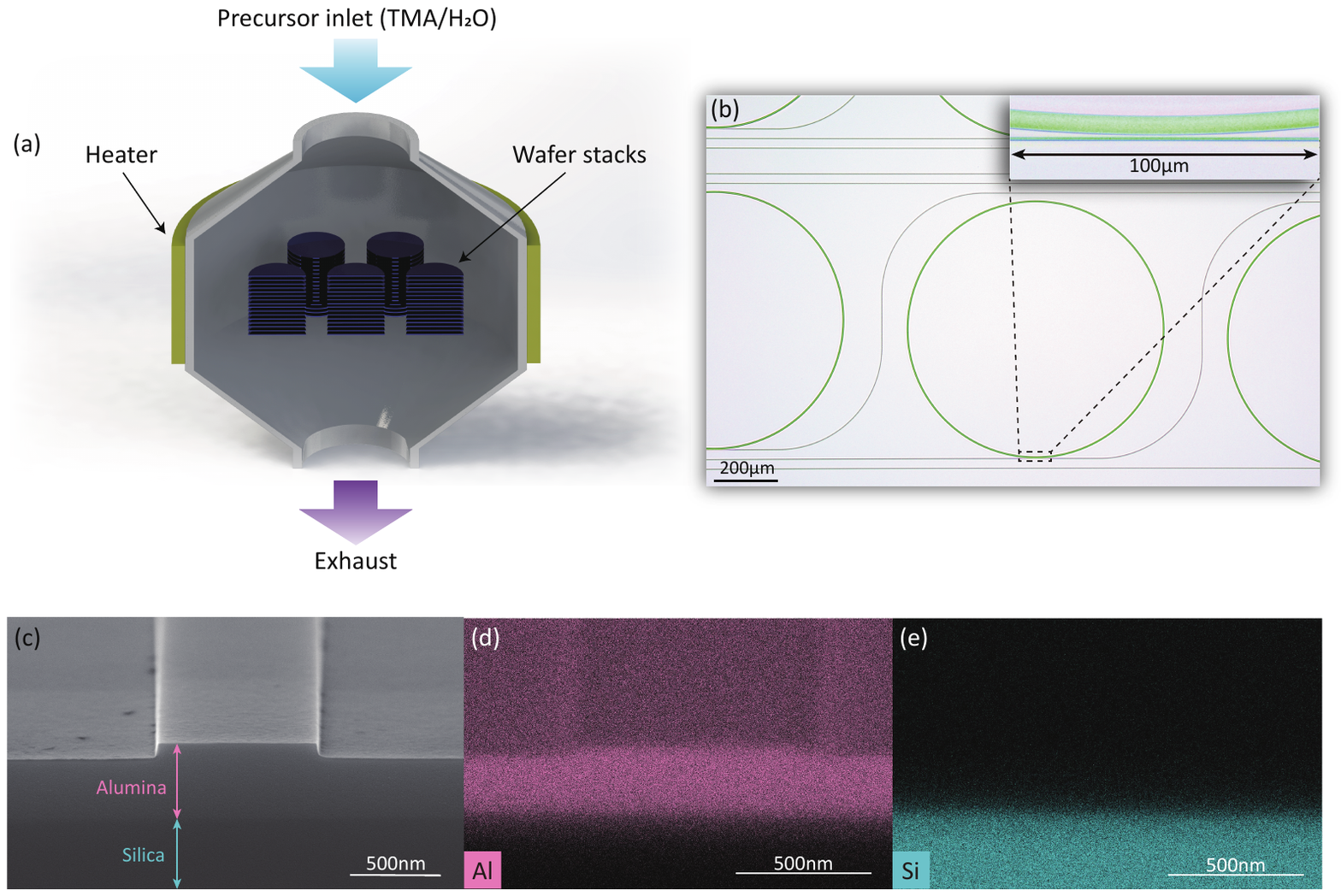}
 \label{fig1}
 \caption{(a) Schematics of ALD chamber capable of processing hundreds of 4" wafers at the same time. (b) Microscope image of patterned ring resonators and interconnection waveguides. Inset: Bus waveguide and ring resonator coupling region. (c) SEM image of cleaved waveguide facet and EDS analysis revealing the elemental mapping of (d) Al (e) Si.}
 \end{figure}
\
\section{Device fabrication and results}
The fabrication started with 4\,$\muup$m of wet thermal oxide grown on silicon wafers. The test wafers were then coated at Entegris by applying a blanket layer of atomic layer deposition (ALD) amorphous alumina. The ALD deposition of alumina coating was performed by sequentially cycling of TMA / H$_2$O with pulsing times between 0.05\,s and 0.15\,s and nitrogen purge times between 18\,s and 20\,s at temperatures between 180\,$^{\circ}$C and 250\,$^{\circ}$C, using a 20” diameter crossflow thermal ALD coatings system custom-built at Entegris in a class 10,000 clean room. The growth rate of this deposition recipe is approximately 1.1$\,$\r{A}$\,$/$\,$cycle. The thickness of the alumina coatings was measured to be a nominal 420\,nm, determined by spectroscopic reflectometry using an Angstrom Sun SR300 system. The ring resonators and associated bus waveguides were defined with a 100kV electron-beam lithography system (Raith EBPG 5200+) with a negative FOx-16 resist. To mitigate electron charging effects due to the highly insulating alumina and silica layers, 200\,nm of poly(4-styrene sulfonic acid) (PSSA) was spun on top of FOx-16 resist before 10$\,$nm of gold was sputtered to provide grounding for stray electrons. The PSSA is water soluble to help the removal of gold after e-beam lithography. After the removal of gold by dipping in water, the chip was developed in 25 percent tetramethylammonium hydroxide (TMAH) developer. The pattern was then transferred to the alumina layer using an Oxford PlasmaPro 100 Cobra Inductively Coupled Plasma Reactive Ion Etching (ICP-RIE) system with a BCl$_3$ based etching recipe. Leftover FOx-16 resist was then removed from the chip by dipping the chip in 10:1 buffered oxide etch for 10 seconds. To further reduce the absorptive loss in alumina waveguide, the chip was annealed in the atmosphere at 500\,$^\circ$C (for 390nm chip) and 600\,$^\circ$C (for 488.5nm chip) for 5 hours to achieve the lowest loss and while avoiding crystallization, which has been reported to take place above 800$^\circ$C\cite{GaNalu,aluanneal}. 

To characterize the ring resonators, we construct a sweeping blue/UV laser by frequency doubling a Ti-Sapphire laser (M2 SolsTiS, 700-1000\,nm) to 390\,nm and 488.5\,nm. The Ti-Sapphire laser is locked to an external cavity, ensuring <50kHz linewidth and the wavelength of the laser is precisely determined by a 0.1\,pm resolution wavemeter, or 0.05\,pm resolution after frequency doubling. To create a sweeping $\sim$390\,nm laser, the $\sim$780\,nm pump laser from Ti:Sapphire laser is coupled to a lithium triborate (LBO) doubling crystal in a resonant cavity (M2 ECD-X). To create a sweeping $\sim$488.5\,nm laser, $\sim$977\,nm pump laser from Ti:Sapphire laser is sent through a Magnesium-doped Periodically Poled Lithium Niobate (MgO:PPLN, Covesion MSHG 976-0.5-30) crystal to frequency double to ~488.5\,nm. The MgO:PPLN crystal is put in an oven, whose temperature is adjusted as the frequency of the pump laser is scanned to maintained phase matching condition of the MgO:PPLN crystal for maximal frequency doubling efficiency. It should be noted that the output power from the Ti:Sapphire laser is wavelength dependent as the spacing of the \'etalon in the resonant cavity is being continuously tuned. The extended transmittance spectrum covering two FSRs is therefore stitched from four continuous scans around 390\,nm (five around 488nm), with the \'etalon spacing being retuned for maximum power output at the beginning of each piecewise scan.% Amorphous alumina is reported to experience phase change to become poly-crystalline gamma phase alumina above 800$^\circ$C\cite{GaNalu}. However, in our case, at an annealing temperature as low as 650$^\circ$C, we start to observe a significant increase in the loss while SEM analysis still suggests the morphology of the waveguide to be pristine. Limited by our ability to investigate materials at nanometer scale, here we only provide a hypothesis for the increase of loss at low temperature. Despite maintaining amorphous, at high annealing temperatures below phase change temperature, the barrier for nucleating is nonetheless lowered. This may lead to clustering of alumina, which creates inhomogeneity within the alumina film the size of which is not orders of magnitude smaller than the propagating wavelength, and thus lead to scattering loss. [input from Shu] Further investigations in the mechanisms of increased loss at temperatures below phase change may be carried out by analyzing amorphous alumina film annealed at different temperatures under TEM or other apparatus that have sub-nanometer resolution. 

Fig. 2 shows the transmittance of the alumina ring resonator at $\sim$\,390\,nm and $\sim$\,488.5\,nm, respectively. For the extended transmittance spectrum, the pump Ti:Sapphire laser is scanned at 500\,MHz/s, corresponding to a scan speed of 1\,GHz/s after frequency doubling. For the zoomed in resonances depicted in the insets, to ensure high wavelength resolution, the pump Ti:Sapphire laser is scanned at 200\,MHz/s (400\,MHz/s after frequency doubling). Multiple sets of resonance peaks can be observed for both the 390\,nm and 488.5\,nm cases, as the 4.5\,$\muup$m wide waveguide supports multiple TE transmission modes. Despite this, the coupling to TE00 mode is being optimized while the coupling to other modes are suppressed. For TE00 modes at ~390\,nm, the TE00 modes exhibit an FSR of 65.6\,GHz, while the TE10 modes have an FSR of 66.3\,GHz, as predicted by the 400\,$\muup$m radius ring geometry. Even higher TE modes are not prominent. One of the TE00 mode resonance peaks demonstrates a loaded Q factor of 1.2\,M, and has an extinction ratio of 2.4\,dB. Using the formula $Q_\mathrm{int}=\frac{2Q_\mathrm{L}}{1+\sqrt{10^{-\mathrm{ER}/10}}}$ (Here $Q_\mathrm{int}$ stands for intrinsic Q, $Q_\mathrm{L}$ stands for loaded Q, and ER stands for extinction ratio in dB.), for under-coupled conditions, we obtain an intrinsic Q of 1.5\,M for this resonance. For TE modes at $\sim$\,488.5\,nm, the FSR is 68.5\,GHz for TE00 mode and 68.6\,GHz for TE10 modes, with one of the TE00 mode resonance peaks demonstrating a high loaded Q of 1.4\,M and has an extinction ratio of 3.2\,dB, corresponding to a loaded Q of 1.9\,M. 

The current Q of our device is likely limited by the residual absorption of alumina and the scattering of the remaining alumina sidewall roughness as the radiation loss limited Q for the resonator is calculated to be beyond 10$^{10}$ for both TE00 and TE10 modes at wavelengths shorter than 500\,nm. Since the modal absorption for TE00 and TE10 mode is the same, the Q difference between the two sets of resonances can be attributed to coupling loss and scattering loss. The waveguide is also capable of transmitting TM modes, however, the confinement of bus waveguide is weak for TM modes and the radiation loss limited Q for TM modes of ring resonator is calculated to be <10\,M. Thus, we did not perform any further measurements of TM mode transmittance. 

In Fig.~3, we compare the performance of our alumina ring resonator to other recent works on UV and blue band photonics. At wavelengths larger than \,450\,nm, low confinement Si$_3$N$_4$ ring resonators with >\,1.5\,mm radii still hold the record for quality factors\cite{5M,6MAMO}. At shorter wavelengths, absorption inherent to the Si$_3$N$_4$ waveguide core would drastically impacts the performance of these devices. AlN was the star material for nanophotonic devices operating at UV-blue band, and progress in AlN film quality boosted the quality factor of AlN based resonators to up to 2.1$\,\times\,$10$^5$ at 390\,nm\cite{AlN}. Alumina film deposited with reactive sputtering and ALD boasts even larger bandgap compared to AlN and renewed the quality factor record to 4.7$\,\times\,$10$^5$ at 405\,nm\cite{twente,aluwgd}. With ALD deposited alumina film and optimized geometry, our alumina ring resonator raises the quality factor record at UV band once again to 1.5$\,\times\,$10$^6$ at 390\,nm.

\begin{figure*}[h!]
\centering
\includegraphics[width=0.70\textwidth]{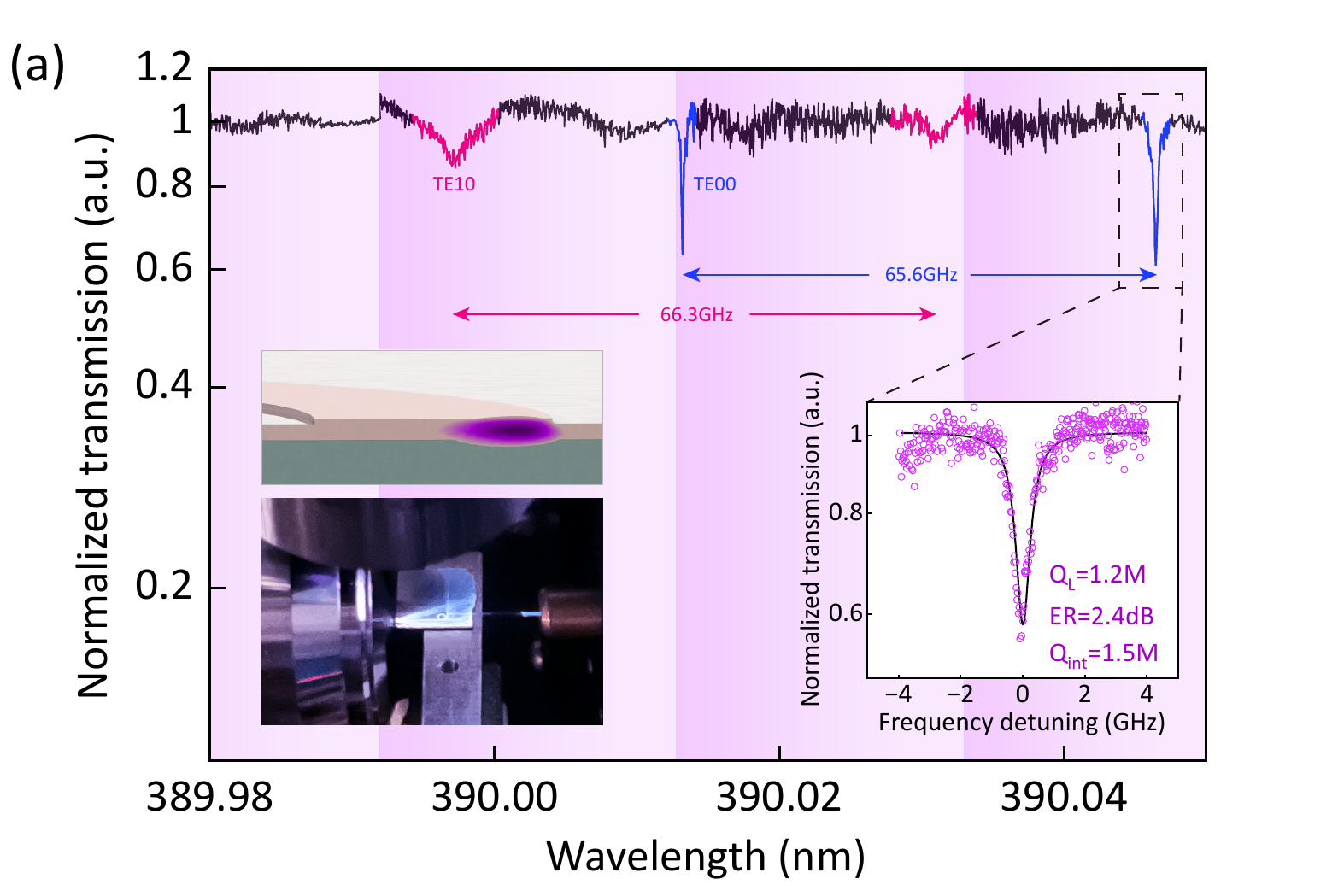}
\includegraphics[width=0.70\textwidth]{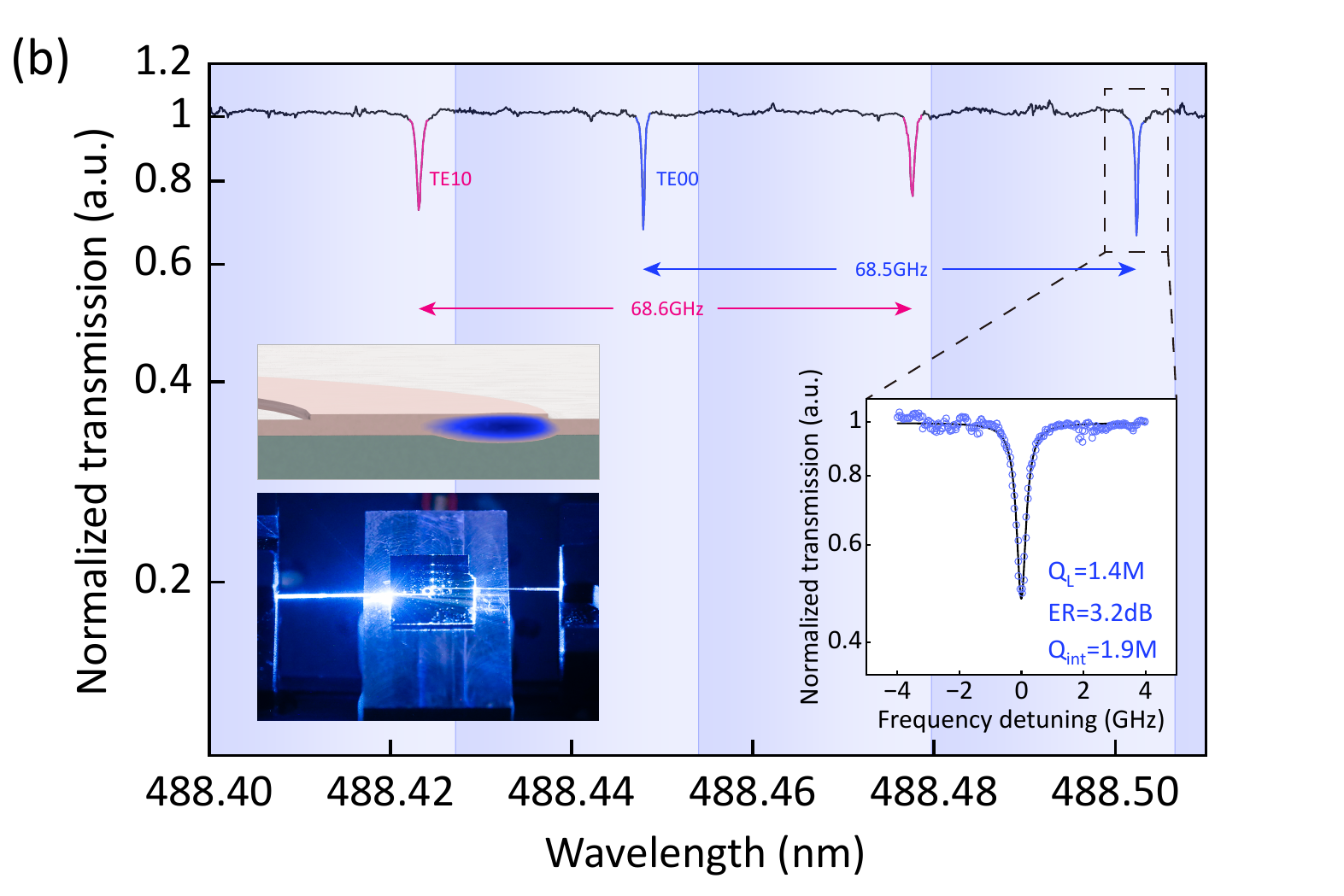}
\label{fig2}
\caption{Transmittance spectrum of alumina ring resonators at (a) 390\,nm and (b) 488.5\,nm showing two sets of transmitted TE modes. Insets top left: Simulated TE00 propagation modes in ring resonators at 390\,nm and 488.5\,nm respectively. Insets bottom left: Ring resonators under test at 390\,nm and 488.5\,nm respectively. Insets right: Zoomed-in views of TE00 resonances demonstrating the highest loaded and intrinsic Q of 1.2\,M\,/\,1.5\,M at 390\,nm and 1.4\,M\,/\,1.9\,M at 488.5\,nm respectively. $Q_\mathrm{L}$ stands for loaded Q, ER stands for extinction ratio in dB and $Q_\mathrm{int}$ stands for intrinsic Q.}
\end{figure*}

\begin{figure*}[h!]
\centering\includegraphics[width=0.85\textwidth]{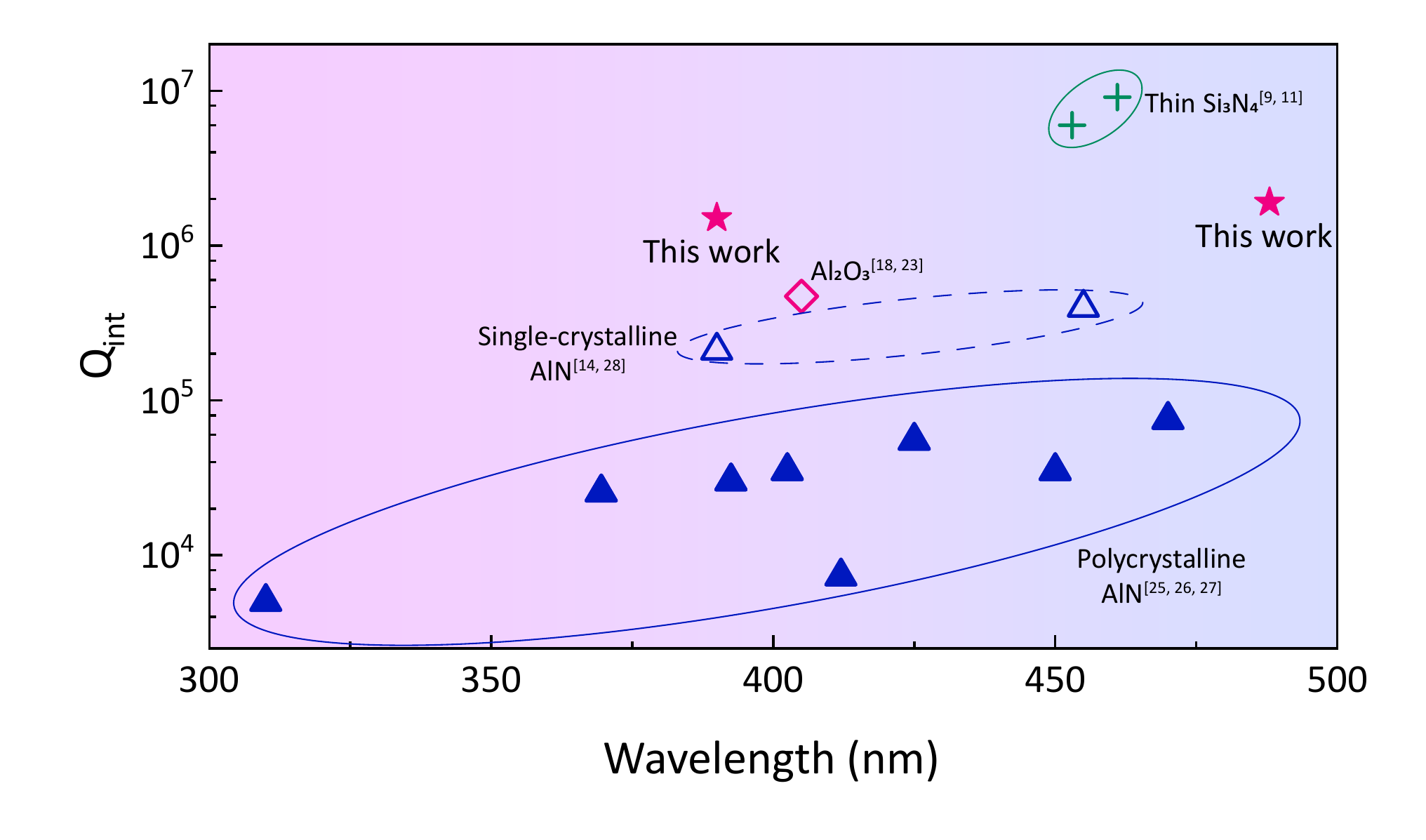}
\label{record}
\caption{Comparison of $Q_{\mathrm{int}}$ vs. wavelength (300\,-\,500\,nm) for recent reports on nanophotonic devices using different materials and geometries. At >\,450\,nm wavelengths, low confinement Si$_3$N$_4$ ring resonators with >\,1.5\,mm radius still hold the record for quality factors\cite{5M,6MAMO}. At shorter wavelengths, improvements in waveguide core materials from polycrystalline AlN\cite{AlNuring2,AlNudisk,AlNcavity3} to single-crystalline AlN\cite{AlN,AlNuring1} to alumina\cite{aluwgd,twente} allow for a boost in quality factor records.}
\end{figure*}

From the Q measurements, the propagation loss of the current ring resonator can be derived to be 0.84\,dB/cm at 390\,nm and 0.51\,dB/cm at 488.5\,nm based on the expression $\alpha = 4.343\times\frac{2\pi n_g}{Q_{\mathrm{int}}\lambda}$, where $n_g$ is the group refractive index obtained through $n_g=\frac{c}{2\pi R_{\mathrm{ring}}*\mathrm{FSR}}$. The superior performance of the ring resonator in this paper can be attributed to the implementation of low-loss amorphous alumina as the waveguide's core material, as well as the ring geometry, which utilizes shallow etching to reduce the scattering loss.

\section{Conclusion}

In conclusion, we demonstrate ultra-high-Q UV and blue band ring resonators featuring low loss ALD alumina as the waveguide core, and optimized geometry which sees propagation mode strongly confined within the alumina waveguide core. This work pushes the intrinsic Q record of high confinement ring resonators to 1.5\,M at 390\,nm and 1.9\,M at 488.5\,nm, corresponding to a propagation loss of only 0.84\,dB/cm and 0.51\,dB/cm respectively. Our results present an important solution in terms of material choice and waveguide design to achieve low-loss integrated photonics in UV and blue band.

\section*{}
\begin{backmatter}
\bmsection{Funding}
This work is funded in part by the Office of Naval Research (ONR) grant N00014-20-1-2693. The materials used in this work is developed under the support of Department of Energy under grant No. DE-SC0019406.  

\bmsection{Acknowledgments}
The authors thanks Michael Rooks, Yong Sun, Lauren McCabe and Kelly Woods for support in the cleanroom and assistance in device fabrication. 

\bmsection{Disclosures}
The authors declare no conflicts of interest.

\bmsection{Data availability} Data is available upon reasonable request. 
\end{backmatter}

%%%%%%%%%% If using BibTeX:
\bibliography{references}

\begin{thebibliography}{10}
\newcommand{\enquote}[1]{``#1''}

\bibitem{AMOARC}
Z.~L. Newman, V.~Maurice, T.~Drake, J.~R. Stone, T.~C. Briles, D.~T. Spencer,
  C.~Fredrick, Q.~Li, D.~Westly, B.~R. Ilic, B.~Shen, M.-G. Suh, K.~Y. Yang,
  C.~Johnson, D.~M.~S. Johnson, L.~Hollberg, K.~J. Vahala, K.~Srinivasan, S.~A.
  Diddams, J.~Kitching, S.~B. Papp, and M.~T. Hummon, \enquote{{Architecture
  for the photonic integration of an optical atomic clock},}
  {\protect\JournalTitle{Optica}} \textbf{6}, 680--685 (2019).

\bibitem{biosensing}
M.~Nissen, B.~Doherty, J.~Hamperl, J.~Kobelke, K.~Weber, T.~Henkel, and M.~A.
  Schmidt, \enquote{{UV Absorption Spectroscopy in Water-Filled Antiresonant
  Hollow Core Fibers for Pharmaceutical Detection},}
  {\protect\JournalTitle{Sensors}} \textbf{18} (2018).

\bibitem{biosensing2}
A.~B.~T. Ghisaidoobe and S.~J. Chung, \enquote{{Intrinsic Tryptophan
  Fluorescence in the Detection and Analysis of Proteins: A Focus on F\"orster
  Resonance Energy Transfer Techniques},} {\protect\JournalTitle{International
  Journal of Molecular Sciences}} \textbf{15}, 22518--22538 (2014).

\bibitem{LiFi}
H.~Haas, L.~Yin, Y.~Wang, and C.~Chen, \enquote{{What is LiFi?}}
  {\protect\JournalTitle{Journal of Lightwave Technology}} \textbf{34},
  1533--1544 (2016).

\bibitem{atomsensing}
J.~Ye, H.~J. Kimble, and H.~Katori, \enquote{{Quantum State Engineering and
  Precision Metrology Using State-Insensitive Light Traps},}
  {\protect\JournalTitle{Science}} \textbf{320}, 1734--1738 (2008).

\bibitem{Rydsensing}
D.~H. Meyer, Z.~A. Castillo, K.~C. Cox, and P.~D. Kunz, \enquote{{Assessment of
  Rydberg atoms for wideband electric field sensing},}
  {\protect\JournalTitle{Journal of Physics B: Atomic, Molecular and Optical
  Physics}} \textbf{53}, 034001 (2020).

\bibitem{mehta2020integrated}
K.~K. Mehta, C.~Zhang, M.~Malinowski, T.-L. Nguyen, M.~Stadler, and J.~P. Home,
  \enquote{{Integrated optical multi-ion quantum logic},}
  {\protect\JournalTitle{Nature}} \textbf{586}, 533--537 (2020).

\bibitem{niffenegger2020integrated}
R.~J. Niffenegger, J.~Stuart, C.~Sorace-Agaskar, D.~Kharas, S.~Bramhavar, C.~D.
  Bruzewicz, W.~Loh, R.~T. Maxson, R.~McConnell, D.~Reens, G.~N. West, J.~M.
  Sage, and J.~Chiaverini, \enquote{{Integrated multi-wavelength control of an
  ion qubit},} {\protect\JournalTitle{Nature}} \textbf{586}, 538--542 (2020).

\bibitem{6MAMO}
N.~Chauhan, J.~Wang, D.~Bose, K.~Liu, R.~L. Compton, C.~Fertig, C.~W. Hoyt, and
  D.~J. Blumenthal, \enquote{{Ultra-low loss visible light waveguides for
  integrated atomic, molecular, and quantum photonics},}
  {\protect\JournalTitle{Opt. Express}} \textbf{30}, 6960--6969 (2022).

\bibitem{AMO780}
Y.-H. Lai, D.~Eliyahu, S.~Ganji, R.~Moss, I.~Solomatine, E.~Lopez, E.~Tran,
  A.~Savchenkov, A.~Matsko, and S.~Williams, \enquote{{780 nm narrow-linewidth
  self-injection-locked WGM lasers},} in \emph{Laser Resonators,
  Microresonators, and Beam Control XXII,}  vol. 11266 A.~V. Kudryashov, A.~H.
  Paxton, V.~S. Ilchenko, and A.~M. Armani, eds., International Society for
  Optics and Photonics (SPIE, 2020), p. 112660O.

\bibitem{5M}
T.~J. Morin, L.~Chang, W.~Jin, C.~Li, J.~Guo, H.~Park, M.~A. Tran,
  T.~Komljenovic, and J.~E. Bowers, \enquote{{CMOS-foundry-based blue and
  violet photonics},} {\protect\JournalTitle{Optica}} \textbf{8}, 755--756
  (2021).

\bibitem{hugerange}
A.~Siddharth, T.~Wunderer, G.~Lihachev, A.~S. Voloshin, C.~Haller, R.~N. Wang,
  M.~Teepe, Z.~Yang, J.~Liu, J.~Riemensberger, N.~Grandjean, N.~Johnson, and
  T.~J. Kippenberg, \enquote{Near ultraviolet photonic integrated lasers based
  on silicon nitride,} {\protect\JournalTitle{APL Photonics}} \textbf{7},
  046108 (2022).

\bibitem{SiN410}
M.~Corato-Zanarella, A.~Gil-Molina, X.~Ji, M.~C. Shin, A.~Mohanty, and
  M.~Lipson, \enquote{{Widely tunable and narrow-linewidth chip-scale lasers
  from near-ultraviolet to near-infrared wavelengths},}
  {\protect\JournalTitle{Nature Photonics}} \textbf{17}, 157--164 (2023).

\bibitem{AlN}
X.~Liu, A.~W. Bruch, Z.~Gong, J.~Lu, J.~B. Surya, L.~Zhang, J.~Wang, J.~Yan,
  and H.~X. Tang, \enquote{{Ultra-high-Q UV microring resonators based on a
  single-crystalline AlN platform},} {\protect\JournalTitle{Optica}}
  \textbf{5}, 1279--1282 (2018).

\bibitem{alumina1}
S.~Toyoda, T.~Shinohara, H.~Kumigashira, M.~Oshima, and Y.~Kato,
  \enquote{{Significant increase in conduction band discontinuity due to solid
  phase epitaxy of Al$_2$O$_3$ gate insulator films on GaN semiconductor},}
  {\protect\JournalTitle{Applied Physics Letters}} \textbf{101}, 231607 (2012).

\bibitem{alumina2}
E.~O. Filatova and A.~S. Konashuk, \enquote{{Interpretation of the Changing the
  Band Gap of Al$_2$O$_3$ Depending on Its Crystalline Form: Connection with
  Different Local Symmetries},} {\protect\JournalTitle{The Journal of Physical
  Chemistry C}} \textbf{119}, 20755--20761 (2015).

\bibitem{sapphire}
J.~Robertson, \enquote{{Band offsets of wide-band-gap oxides and implications
  for future electronic devices},} {\protect\JournalTitle{Journal of Vacuum
  Science \& Technology B: Microelectronics and Nanometer Structures
  Processing, Measurement, and Phenomena}} \textbf{18}, 1785--1791 (2000).

\bibitem{aluwgd}
G.~N. West, W.~Loh, D.~Kharas, C.~Sorace-Agaskar, K.~K. Mehta, J.~Sage,
  J.~Chiaverini, and R.~J. Ram, \enquote{{Low-loss integrated photonics for the
  blue and ultraviolet regime},} {\protect\JournalTitle{APL Photonics}}
  \textbf{4}, 026101 (2019).

\bibitem{aluwgd2}
M.~M. Aslan, N.~A. Webster, C.~L. Byard, M.~B. Pereira, C.~M. Hayes, R.~S.
  Wiederkehr, and S.~B. Mendes, \enquote{{Low-loss optical waveguides for the
  near ultra-violet and visible spectral regions with Al$_2$O$_3$ thin films
  from atomic layer deposition},} {\protect\JournalTitle{Thin Solid Films}}
  \textbf{518}, 4935--4940 (2010).

\bibitem{activealufilm}
R.~Wang, H.~C. Frankis, H.~M. Mbonde, D.~B. Bonneville, and J.~D. Bradley,
  \enquote{{Erbium-ytterbium co-doped aluminum oxide thin films: Co-sputtering
  deposition, photoluminescence, luminescent lifetime, energy transfer and
  quenching fraction},} {\protect\JournalTitle{Optical Materials}}
  \textbf{111}, 110692 (2021).

\bibitem{activealu}
C.~I. van Emmerik, M.~Dijkstra, M.~de~Goede, L.~Chang, J.~Mu, and S.~M.
  Garcia-Blanco, \enquote{{Single-layer active-passive Al$_2$O$_3$ photonic
  integration platform},} {\protect\JournalTitle{Opt. Mater. Express}}
  \textbf{8}, 3049--3054 (2018).

\bibitem{SiNAlu}
Z.~Su, N.~Li, H.~C. Frankis, E.~S. Magden, T.~N. Adam, G.~Leake, D.~Coolbaugh,
  J.~D.~B. Bradley, and M.~R. Watts, \enquote{{High-Q-factor Al$_2$O$_3$
  micro-trench cavities integrated with silicon nitride waveguides on
  silicon},} {\protect\JournalTitle{Opt. Express}} \textbf{26}, 11161--11170
  (2018).

\bibitem{twente}
C.~Franken, W.~Hendriks, L.~Winkler, M.~Dijkstra, A.~do~Nascimento~Jr., A.~van
  Rees, M.~Mardani, R.~Dekker, J.~van Kerkhof, P.~van~der Slot,
  S.~Garc\'ia-Blanco, and K.-J. Boller, \enquote{{Hybrid integrated near UV
  lasers using the deep-UV Al$_2$O$_3$ platform},}  (2023).

\bibitem{GaNalu}
J.~Jang, \enquote{Study on cavity engineered sapphire substrate for highly
  efficient gan-based light-emitting diodes,} Ph.D. thesis, Seoul National
  University (2018).

\bibitem{aluanneal}
V.~V. Afanas’ev, A.~Stesmans, B.~J. Mrstik, and C.~Zhao, \enquote{Impact of
  annealing-induced compaction on electronic properties of
  atomic-layer-deposited {Al$_2$O$_3$},} {\protect\JournalTitle{Applied Physics
  Letters}} \textbf{81}, 1678--1680 (2002).

\bibitem{AlNuring2}
T.-J. Lu, M.~Fanto, H.~Choi, P.~Thomas, J.~Steidle, S.~Mouradian, W.~Kong,
  D.~Zhu, H.~Moon, K.~Berggren, J.~Kim, M.~Soltani, S.~Preble, and D.~Englund,
  \enquote{{Aluminum nitride integrated photonics platform for the ultraviolet
  to visible spectrum},} {\protect\JournalTitle{Opt. Express}} \textbf{26},
  11147--11160 (2018).

\bibitem{AlNudisk}
M.~Bürger, G.~Callsen, T.~Kure, A.~Hoffmann, A.~Pawlis, D.~Reuter, and D.~J.
  As, \enquote{Lasing properties of non-polar {GaN} quantum dots in cubic
  aluminum nitride microdisk cavities,} {\protect\JournalTitle{Applied Physics
  Letters}} \textbf{103}, 021107 (2013).

\bibitem{AlNcavity3}
S.~Sergent, M.~Arita, S.~Kako, K.~Tanabe, S.~Iwamoto, and Y.~Arakawa,
  \enquote{High-{Q} {AlN} photonic crystal nanobeam cavities fabricated by
  layer transfer,} {\protect\JournalTitle{Applied Physics Letters}}
  \textbf{101}, 101106 (2012).

\bibitem{AlNuring1}
W.~Shin, Y.~Sun, M.~Soltani, and Z.~Mi, \enquote{{Demonstration of green and UV
  wavelength high Q aluminum nitride on sapphire microring resonators
  integrated with microheaters},} {\protect\JournalTitle{Applied Physics
  Letters}} \textbf{118}, 211103 (2021).

\end{thebibliography}

%%%%%%%%%% If preparing manually:
% \newcommand{\enquote}[1]{``#1''}

% \bibitem{Zhang:14}
% Y.~Zhang, S.~Qiao, L.~Sun, Q.~W. Shi, W.~Huang, L.~Li, and Z.~Yang,
%   \enquote{Photoinduced active terahertz metamaterials with nanostructured
%   vanadium dioxide film deposited by sol-gel method,}
%   {\protect\JournalTitle{Optics Express}} \textbf{22}, 11070--11078 (2014).

% \bibitem{Optica}
% {Optica}, \enquote{{Optica Publishing Group},}
%   \url{http://www.opg.optica.org}.

% \bibitem{FORSTER2007}
% P.~Forster, V.~Ramaswamy, P.~Artaxo, T.~Bernsten, R.~Betts, D.~Fahey,
%   J.~Haywood, J.~Lean, D.~Lowe, G.~Myhre, J.~Nganga, R.~Prinn, G.~Raga,
%   M.~Schulz, and R.~V. Dorland, \enquote{Changes in atmospheric consituents and
%   in radiative forcing,} in \enquote{Climate Change 2007: The Physical Science
%   Basis. Contribution of Working Group 1 to the Fourth assesment report of
%   Intergovernmental Panel on Climate Change,}  S.~Solomon, D.~Qin, M.~Manning,
%   Z.~Chen, M.~Marquis, K.~B. Averyt, M.~Tignor, and H.~L. Miler, eds.
%   (Cambridge University Press, 2007).

% \end{thebibliography}

\end{document}